\documentclass[twocolumn,floatfix,preprintnumbers]{revtex4-1}

\usepackage[utf8]{inputenc}
\usepackage[colorlinks=true,citecolor=blue,linkcolor=blue]{hyperref}
\usepackage[normalem]{ulem}
\usepackage{url}
\usepackage{graphicx,wrapfig,float,slashed,cancel}
\usepackage{amsmath,amssymb,epsfig,graphicx,xcolor,stmaryrd}
\usepackage{physics}
\usepackage{bm}
\usepackage{enumitem}

\definecolor{darkblue}{RGB}{1, 90, 173}

\begin{document}


\title{Modifications on parameters of  $Z(4430)$  in a dense medium}

\author{K. Azizi}
\email{kazem.azizi@ut.ac.ir}
\affiliation{Department of Physics, University of Tehran, North Karegar Ave. Tehran 14395-547, Iran}
\affiliation{ Department of Physics, Do\v{g}u\c{s} University, Acibadem-Kadik\"{o}y, 34722
Istanbul, Turkey}

\author{N. Er}
\email{nuray@ibu.edu.tr}
\affiliation{ Department of Physics, Bolu  Abant \.{I}zzet Baysal University,
G\"olk\"oy Kamp\"us\"u, 14980 Bolu, Turkey}

\date{\today}

\preprint{}

\begin{abstract}
The charmonium-like resonance $ Z_c(3900) $ and its excited state $ Z(4430) $ are among the particles that are  serious candidates for double heavy tetraquarks. Calculations of different  parameters associated with  these states both in the vacuum and the medium with finite density are of great importance. Such investigations help us clarify their nature, internal quark-gluon organization and quantum numbers. In this accordance,  we extend our previous analyses on the ground state $ Z_c(3900) $ to investigate the medium modifications on different parameters of the excited $ Z(4430) $  state. In particular, we calculate the mass, vector self-energy and current coupling of $ Z(4430) $ in terms of density, up to a density comparable to the density of the cores of massive neutron stars. The obtained results may help experimental groups aiming to study the behavior of exotic states at higher densities. 
\end{abstract}


\maketitle

\section{Introduction} \label{sec:intro}
Over the past two decades, the results of many experimental observations on some resonances have shown that these states can not be put in the class of the standard  mesons and  baryons since their spectrum and decay patterns differ from the standard hadrons, considerably. These results have led us to collect  these states under the new title:   the exotic states made of multiquarks, antiquarks and  valence gluons.  Among these states, the tetraquarks have been relatively more  in the focus of the experimental and theoretical studies. The charmonium-like resonances made of a $ c\bar{c} $  and a light $ q_1\bar{q_2} $ pair receive much attentions as they demonstrate different properties.  Investigation of  these states can help us not only clarify their nature and internal structure  but also get useful knowledge on the nature of strong interaction inside these particles. 

The observation of the ground states $Z^{\pm}_c(3900)$ were reported simultaneously by the BESIII  \cite{Ablikim:2013mio}  and  Belle \cite{Liu:2013dau}  Collaborations in 2013. Many aspects of these states have been previously  investigated. For details see for instance Refs. \cite{Agaev:2020zad,Agaev:2016dev,PhysRevD.96.034026,Azizi:2020itk,Ozdem:2017jqh}  and references therein. Hence, we won't spend time on this state in the present work and will focus on the state $ Z(4430) $.  For a first time in 2008, the Belle Collaboration reported a  distinct peak in the $\pi^{\pm}\psi'$ invariant mass distribution  in $B \rightarrow K\pi^{\pm}\psi'$ decay with the statistical significance of $6.5\sigma$. The measured mass and width were $M=4433\pm4(\textrm{stat})\pm2(\textrm{syst})$ MeV and  $\Gamma=45^{+18}_{-13}(\textrm{stat})^{30}_{-13}(\textrm{syst})$ MeV \cite{PhysRevLett.100.142001}. In an amplitude analysis of $B^0 \rightarrow \psi'K^+\pi^-$ decays \cite{PhysRevD.88.074026}, the quantum numbers of the $Z(4430)$ were set to $J^P=1^+$ and the results for the mass and width were obtained as $4485^{+22+28}_{-22-11}$ MeV/${c^2}$ and $200^{+41+26}_{-46-35}$ MeV, respectively. The Belle Collaboration then \cite{PhysRevD.90.112009}, during the observation of a new  charmonium like state $Z^+_c(4200)$, in addition,  found evidence for $Z^+(4430) \rightarrow J/\psi\pi^+$.  In the first independent confirmation by LHCb Collaboration \cite{PhysRevLett.112.222002}, its spin-parity was assigned as  $1^+$. The  mass of the resonance, $4475\pm7^{+15}_{-25}$ MeV, and its width, $172\pm13^{+37}_{-34}$ MeV, were measured.  In the same mass region, an alternative and model-independent confirmation of the existence of a $\psi(2S)\pi$ resonance is later  presented by the LHCb, as well   \cite{PhysRevD.92.112009}. 

In  theoretical side,  various assignments have been made for the nature and quark-gluon structure  of  $Z(4430)$ resonance. Using  different phenomenological and computational models, the mass and some other quantities of this state  have been studied. Considering a molecular structure, the $Z(4430)$ state was studied in \cite{Liu:2007bf,Liu:2008xz,Lee:2007gs,Lee:2008tz,Braaten:2007xw,Zhang:2009vs,Branz:2010sh,Close:2010wq,Wang:2018pwi}. The compact tetraquark or diquark-antidiquark structure  was assigned for this state in Refs.  \cite{Wang:2019hnw,Wang:2019tlw,PhysRevD.96.034026,Maiani:2014aja,Goerke:2016hxf,Wang:2014vha,Bracco:2008jj,Ebert:2008kb,Liu:2008qx} to calculate various observables associated with the $Z(4430)$ resonance. Other theoretical interpretations can be collected as cusp effects \cite{Bugg,Swanson:2014tra,Ikeda} and  hadrocharmonium  \cite{Danilkin:2011sh,Dubynskiy:2008mq}. For more information, one can see  Refs. \cite{Albuquerque:2018jkn,Olsen:2017bmm,Guo:2017jvc,Esposito:2016noz,Lebed:2016hpi,Ali:2017jda,Chen:2016qju,Liu:2019zoy,Hosaka:2016pey}.

 In the present work, we  shall adopt the compact tetraquark/diquark-antidiquark interpretation of the state $Z(4430)$ and consider it as the radial excitation of the ground state $Z_c(3900)$ with the same spin-parity.  In Ref. \cite{Maiani:2014aja}, the authors have made the hypothesis that the spacing (mass difference) in radial excitations in these channels could closely resemble those observed in the  standard P-wave charmonia. Therefore, the authors have considered the state $Z(4430)$ as  the first radial excitation of the $Z_c(3900)$ state. The same scenario has been applied  in different studies  \cite{Wang:2014vha,PhysRevD.96.034026,Chen:2019osl,Zhu:2016arf}, as well.  We calculate the mass and current coupling of $Z(4430)$  state  in the medium with high density, comparable with the density of the neutrons star cores. We discuss the modifications on the considered physical quantities due to the nuclear dense medium. The medium effects are represented by various operators that enter the nonperturbative parts of the calculations (for details see, for instance,  Ref. \cite{Azizi:2014yea} and references therein). We calculate the values of the mass and current coupling at saturation density. These quantities are also obtained in vacuum setting the density of the medium to zero. We compare the obtained results at zero density limit with the existing theoretical predictions and the  experimental data. 

The high energy heavy ion collision experiments are excellent platforms to produce the  heavy standard hadrons  as well as  the exotic particles.  With upgraded detectors at the Relativistic Heavy Ion Collider (RHIC)  at the Bookhaven National Laboratory (BNL) and the Large Hadron Collider (LHC),  now it is  possible to investigate not only the   ground state particles but also their excited states. Thus, these experiments provide   opportunities to study   exotic hadrons like   $Z_c(3900)$ and its excited state $Z(4430)$. The study on other exotic states  that are either molecular states made of various standard hadrons or compact  objects of quarks and antiquarks are now possible, as well. Due to  a large number of heavy quarks and antiquarks produced in these experiments, many exotic states could be formed. Hence, it will be a good opportunity to study the exotic particles  and their higher states in order to determine their nature and internal quark-gluon structure as well as fix their quantum numbers (for more information see for instance \cite{Cho:2017dcy,Cho:2011ew}).  At RHIC and heavy ion collision experiments at LHC it will be also  possible to investigate different phases of hadronic matter and search for quark-gluon-plasma (QGP) as a new possible phase of matter . Some in-medium experiments like  $ \bar{P}ANDA $ at Facility for Antiproton and Ion Research (FAIR) at GSI and  the Nuclotron-based Ion Collider Facility at JINR (NICA) at Dubna aim to study the parameters of exotic states in terms of density.  These experiments can investigate the the $Z^+(4430)$ even further by switching to studies of the $Z^+(4430)$ in formation mode. Due to the charge of the Z, this is only possible by annihilating the antiprotons on a neutron in a deuterium  or hydrogen gas target. They can study the $Z^+(4430)$ state  in both production and formation experiments \cite{Wiedner:2011mf}.

The paper is organized as follows: In next section the parameters of the $Z(4430)$ is calculated in dense medium. In section 3 the numerical analyses of the physical quantities under study is performed. The last section includes the summary and our concluding notes. 

\section{Parameters of  $Z(4430)$ in cold nuclear medium}
In this section, we aim to calculate the mass and  current coupling as well as the  scalar and vector self-energies of the exotic  $Z(4430)$ (in what follows we denote it by $Z$ )  state as an exited state of the $Z_c(3900)$  ($Z_c$) in cold nuclear medium. To this end, the positively charged $Z^+(4430)$ state  with quark content $\bar{c}cu\bar{d}$ is considered.  But, the parameters of the negatively charged state with  $\bar{c}c\bar{u}d$ content do not change as a result of Chiral limit used in the calculations. In this context, our starting point is to consider the following density dependent two-point correlation function:
\begin{equation}\label{corre1}
\Pi_{\mu\nu}(p)=i\int{d^4 xe^{ip\cdot x}\langle\psi_0|\mathcal{T}[J^{Z_{(c)}}_{\mu}(x)J^{{Z_{(c)}} \dagger}_{\nu}(0)]|\psi_0\rangle},
\end{equation}
where $|\psi_0\rangle$ is the ground state of nuclear matter and $J^{Z_{(c)}}_{\mu}$ is the interpolating current of $Z$/$Z_c$ states that couples to these states, simultaneously.  We should note that in the present study we choose to work at zero momentum limit of the particle's three momentum, $ \mid\bf{p}\mid \rightarrow0 $,  for which the longitudinal and transverse polarizations become degenerate, and there is no way to distinguish these two polarizations.  For non-zero and  finite three momentum in nuclear matter, however, the longitudinal and transverse polarization states are distinguishable and they should be calculated, separately (for more details see for instance \cite{Kim:2019ybi}). At zero three momentum limit, $ p^2=p_0^2 $  and the calculation goes forward in the standard way with the above  single function  (see also \cite{COHEN1995221}).
For $J^P=1^+$, the interpolating current is given by
\begin{eqnarray}\label{ }
J^{Z_{(c)}}_{\mu}(x)&=&\frac{i\epsilon^{abc}\epsilon^{dec}}{\sqrt{2}} \Bigg \{ \Big[u_{a}^T(x) C \gamma_5 c_b(x)\Big] \Big[\bar{d}_{d}(x) \gamma_{\mu} C\bar{c}^T_e(x) \Big]  \nonumber \\
&-& \Big[u_{a}^T(x) C \gamma_{\mu} c_b(x)\Big] \Big[\bar{d}_{d}(x) \gamma_5 C\bar{c}^T_e(x) \Big] \Bigg \},
\end{eqnarray}
where, $a, b, c, d, e$ are color indices and $C$ is the charge conjugation operator. 

We need to derive the in-medium sum rules for the mass $m_{Z}$ and current coupling $f_{Z}$ of the excited state $Z$. For this purpose, the previously obtained expressions for the mass and  current coupling of $Z_c$ (see Ref. \cite{Azizi:2020itk})  are considered as input parameters.  The procedure is done in two steps: First, we derive the sum rules for the in-medium mass $m_{Z_c}$ and in-medium current coupling $f_{Z_c}$ of the ground state exotic $Z_c$. For this, we use the "ground state + continuum" approximation by including the $Z(4430)$ state into the list of "higher resonances" and extract  sum rules and corresponding numerical values for $Z_c$ as discussed in Ref. \cite{Azizi:2020itk}.  At the next step,  we  increase the threshold and adopt the "ground state+radially excited state+continuum" scheme, and perform the required standard calculations. In the second case, the parameters of both the $Z_c$ and $Z$ appear in the expression of the hadronic side.  To extract the parameters of $Z$, we use those of the $Z_c$ as inputs. 

Thus, the  phenomenological side of the correlation function is obtained by saturating it with  the  complete sets of hadronic states with the same quantum numbers as the interpolating current.   
Isolating the ground state  $Z_c$ and radially excited $Z$ resonances, considering the "ground state+radially excited state+continuum" scheme, and integrating over $x$, we obtain the phenomenological side of the correlation function in momentum space as
\begin{eqnarray} \label{had1}
\Pi^{Phe}_{\mu\nu}(p)=&-& \frac{\langle\psi_0|J_{\mu}|Z_c(p)\rangle \langle Z_c(p)|J^{\dagger}_{\nu}|\psi_0\rangle}{p^{*2}-m_{Z_c}^{*2}} \nonumber \\
&- &\frac{\langle\psi_0|J_{\mu}|Z(p)\rangle \langle Z(p)|J^{\dagger}_{\nu}|\psi_0\rangle}{p^{*2}-m_{Z}^{*2}} + ... ,
\end{eqnarray}
where $p^*$ is the in-medium momentum; and $m_{Z_c}$ and $m_Z$ are the masses of $Z_c$ and $Z$ states, respectively. In Eq. (\ref{had1}), contributions arising from higher resonances and continuum are represented by the dots.  The current meson couplings of the states under consideration are expressed in terms of the polarization vectors $\varepsilon_{\mu}$ and $ \tilde{\varepsilon}_{\mu}$, respectively, as
\begin{eqnarray} \label{had2}
\langle\psi_0|J_{\mu} |Z_c(p)\rangle &=& f^{*}_{Z_c} m_{Z_c}^{*} \varepsilon_{\mu}, \nonumber \\
\langle\psi_0|J_{\mu} |Z(p)\rangle &=& f^{*}_{Z} m_{Z}^{*} \tilde{\varepsilon}_{\mu}.
\end{eqnarray}
Using Eq. (\ref{had2}) in Eq. (\ref{had1})  and summing over the polarization vectors, we can construct the new form of the function $\Pi^{Phe}_{\mu\nu}(p)$ as
\begin{eqnarray}
\Pi_{\mu\nu}^{Phe}(p)=&-&\frac{m^{*2}_{Z_c} f^{*2}_{Z_c}}{p^{*2}-m^{*2}_{Z_c}} \Big[-g_{\mu\nu} +\frac{p_{\mu}^{*} p^{*}_{\nu} }{m^{*2}_{Z_c}} \Big] \nonumber \\
&-&\frac{m^{*2}_{Z} f^{*2}_{Z}}{p^{*2}-m^{*2}_{Z}} \Big[-g_{\mu\nu} +\frac{p_{\mu}^{*} p^{*}_{\nu} }{m^{*2}_{Z}} \Big] + ... .
\end{eqnarray}
At this point, we should note that a particle in nuclear medium gain two kinds of self-energies: The scalar self-energy, which is defined as the difference between the in-medium and vacuum masses, $\Sigma_s=m_Z^*-m_Z$, and the vector self-energy that enters  the expression of the in-medium momentum, $p^*_{\nu}=p_{\nu}-\Sigma_{\upsilon}u_{\nu}$,  where $\Sigma_{\upsilon}$ is the vector self-energy and $u_{\nu}$ is the four-velocity of the nuclear medium. Our calculations take place in the rest  frame of the medium,  $u_{\nu}=(1,0)$, and so $p^{\nu}\cdot u_{\nu}=p_0$. Consequently,  the correlation function is written as
\begin{eqnarray} \label{PiPhe}
\Pi_{\mu\nu}^{Phe}(p)=&-&\frac{ f^{*2}_{Z_c}}{p^2-\mu_{Z_c}^2} \Big[-g_{\mu\nu} m^{*2}_{Z_c} +p_{\mu}p_{\nu} \nonumber \\
&-&\Sigma^{Z_c}_{\upsilon}p_{\mu}u_{\nu} -\Sigma^{Z_c}_{\upsilon}p_{\nu}u_{\mu}+\Sigma_{\upsilon}^{Z_c 2} u_{\mu}u_{\nu} \Big] \nonumber \\
&-&\frac{ f^{*2}_{Z}}{p^2-\mu_{Z}^2} \Big[-g_{\mu\nu} m^{*2}_{Z} +p_{\mu}p_{\nu} 
- \Sigma^{Z}_{\upsilon}p_{\mu}u_{\nu} \nonumber \\
&-&\Sigma^{Z}_{\upsilon}p_{\nu}u_{\mu}+\Sigma^{Z 2}_{\upsilon} u_{\mu}u_{\nu} \Big] + ...,\nonumber\\
\end{eqnarray}
where $\mu_{Z_{(c)}}^2=m^{*2}_{Z_{(c)}}-\Sigma^{Z_{(c)} 2}_{\upsilon}+2p_0\Sigma^{Z_{(c)} }_{\upsilon}$. After applying the Borel transformation with respect to the parameter $p^2$ to both sides of Eq. (\ref{PiPhe}), the phenomenological side of the correlation function yields, 
\begin{eqnarray} \label{PiPhe2}
\mathbf{ \Pi}_{\mu\nu}^{Phe}(p)&=&f^{*2}_{Z_c}e^{-\mu_{Z_c}^2/M^2} \Big[-g_{\mu\nu} m^{*2}_{Z_c}
+ p_{\mu}p_{\nu}\nonumber \\
& - & \Sigma^{Z_c}_{\upsilon}p_{\mu}u_{\nu} -\Sigma^{Z_c}_{\upsilon}p_{\nu}u_{\mu}+\Sigma_{\upsilon}^{Z_c 2} u_{\mu}u_{\nu} \Big] \nonumber \\
&+&f^{*2}_{Z}e^{-\mu_{Z}^2/M^2} \Big[-g_{\mu\nu} m^{*2}_{Z}
+ p_{\mu}p_{\nu} - \Sigma^{Z}_{\upsilon}p_{\mu}u_{\nu}\nonumber \\
& -&\Sigma^{Z}_{\upsilon}p_{\nu}u_{\mu}+\Sigma^{Z 2}_{\upsilon} u_{\mu}u_{\nu} \Big] \nonumber \\&+& ...,
\end{eqnarray}
where $M^2$ is the Borel mass parameter. 

The next step is to calculate the QCD side of the correlation function. This is done via the usage of the explicit form of the interpolating current in the correlation function. The quark fields are contracted in the presence of the dense medium via the Wick's theorem. This results in the expression of the correlation function in terms of the in-medium heavy and light propagators:
\begin{widetext}
\begin{eqnarray}\label{qcd}
&&\Pi_{\mu\nu}^{QCD}(p)= -\frac{i}{2}\varepsilon_{abc} \varepsilon_{a'b'c'} \varepsilon_{dec} \varepsilon_{d'e'c'}  \int d^4 x e^{ipx}\Big\{Tr\Big[\gamma_5 \tilde{S}_u^{aa'} (x)\gamma_5 S_c^{bb'}(x)\Big] \nonumber\\
&& Tr\Big[\gamma_{\mu} \tilde{S}_c^{e'e} (-x)\gamma_{\nu} S_d^{d'd}(-x)\Big] - Tr\Big[[\gamma_{\mu} \tilde{S}_c^{e'e} (-x)  \gamma_5 S_d^{d'd}(-x)\Big]Tr\Big[\gamma_{\nu} \tilde{S}_u^{aa'} (x)\gamma_5 S_c^{bb'}(x)\Big] -\nonumber \\
&& Tr\Big[\gamma_5 \tilde{S}_u^{aa'} (x)\gamma_{\mu} S_c^{bb'}(x)\Big] Tr\Big[\gamma_5 \tilde{S}_c^{e'e} (-x)\gamma_{\nu} S_d^{d'd}(-x)\Big]+ \nonumber \\
&& Tr\Big[\gamma_{\nu}  \tilde{S}_u^{aa'} (x)\gamma_{\mu} S_c^{bb'}(x)\Big] Tr\Big[\gamma_5 \tilde{S}_c^{e'e} (-x)\gamma_5 S_d^{d'd}(-x)\Big]  \Big\}_{|\psi_0\rangle},\nonumber \\
\end{eqnarray}
\end{widetext}
where $\tilde{S}_{q(c)}= C S_{q(c)}^{T}C$. The in-medium heavy and light quarks' propagators are used in coordinate space and the calculations are transferred to the momentum space by performing the Fourier integrals. To suppress the contributions of the higher states and continuum the Borel transformations are applied based on the standard prescriptions of the method. To further suppress the contributions of the unwanted states, continuum subtraction supplied by the quark-hadron duality assumption is applied. All of these procedures are described in Ref.  \cite{Azizi:2020itk}.  We should stress that we take into account the non-perturbative operators up to five dimensions in the QCD side. The contributions of the six and upper dimensional operators are expected to be small and they are neglected. Thus, the QCD side of the correlation function in terms of  the selected Lorentz structures is written as
\begin{eqnarray}\label{QCDcof}
&&\Pi_{\mu\nu}^{QCD}(M^2, s_0^*)=\mathbf{ \Upsilon}^{QCD}_{1}(M^2, s_0^*)(-g_{\mu\nu} )\nonumber \\
&+& \mathbf{ \Upsilon}^{QCD}_{2}(M^2, s_0^*) p_{\mu}p_{\nu}   
+\mathbf{ \Upsilon}^{QCD}_{3}(M^2, s_0^*)(- p_{\mu}u_{\nu})   \nonumber \\
&+&\mathbf{ \Upsilon}^{QCD}_{4}(M^2, s_0^*)  (-p_{\nu}u_{\mu} )
+ \mathbf{ \Upsilon}^{QCD}_{5}(M^2, s_0^*) u_{\mu}u_{\nu}.\nonumber \\
\end{eqnarray}
The Borel transformed invariant functions $\mathbf{ \Upsilon}^{QCD}_{i}(M^2, s_0^*)$ in Eq.  (\ref{QCDcof}) are represented in terms of the  two-point spectral densities, $\rho_i^{QCD}(s)$, related to the  imaginary parts of the selected coefficients:
 \begin{equation}\label{Upsil2}
\mathbf{ \Upsilon}^{QCD}_{i}(M^2, s_0^*)= \int_{4m_c^2}^{s^{*}_0} ds \rho^{QCD}_{i}(s)e^{-\frac{s}{M^2}},
\end{equation}
where $s_0^*$ is the in-medium continuum threshold parameter  separating  the contributions of the ground state $Z_c$ and the first excited state $ Z $  from the  higher resonances and continuum.  $ i $ runs from $ 1 $ to $ 5 $. The spectral density corresponding to each structure is expressed in terms of perturbative and non-perturbative contributions as follows: 
\begin{eqnarray}\label{ }
\rho_i^{QCD} (s)&=&\rho_i^{pert}(s) + \rho_i^{qq} (s) + \rho_i^{gg} (s) + \rho_i^{qgq} (s),
\end{eqnarray}
where $qq $, $gg $  and $qgq $ denote the two quark, two gluon and mixed quark-gluon condensates as the non-perturbative effects, respectively. As an example, the spectral density corresponding to the structure $ g_{\mu\nu} $ is given in Ref. \cite{Azizi:2020itk}.

After all these lengthy calculations, when the phenomenological and QCD results of each structure are compensate, the sum rules for the mass, vector self-energy and current coupling constant of $Z$ state are obtained as follows:
\begin{eqnarray}\label{SumR}
m^{*2}_{Z_c}f^{*2}_{Z_c} e^{-\frac{\mu_{Z_c}^2}{M^2}} +m^{*2}_{Z}f^{*2}_{Z} e^{-\frac{\mu_{Z}^2}{M^2}} & = & \mathbf{ \Upsilon}^{QCD}_{1}(M^2, s_0^*) \label{SR1},   \\
f^{*2}_{Z_c} e^{-\frac{\mu_{Z_c}^2}{M^2}} + f^{*2}_{Z} e^{-\frac{\mu_{Z}^2}{M^2}} & = & \mathbf{ \Upsilon}^{QCD}_{2}(M^2, s_0^*) \label{SR2},  \\
\Sigma^{Z_c}_{\upsilon}f^{*2}_{Z_c} e^{-\frac{\mu_{Z_c}^2}{M^2}} +\Sigma^{Z}_{\upsilon}f^{*2}_{Z} e^{-\frac{\mu_{Z}^2}{M^2}}& = &\mathbf{  \Upsilon}^{QCD}_{3}(M^2, s_0^*) \label{SR3},  \\
\Sigma^{Z_c}_{\upsilon}f^{*2}_{Z_c} e^{-\frac{\mu_{Z_c}^2}{M^2}} +\Sigma^{Z}_{\upsilon}f^{*2}_{Z} e^{-\frac{\mu_{Z}^2}{M^2}} & = & \mathbf{ \Upsilon}^{QCD}_{4}(M^2, s_0^*) \label{SR4},   \\
\Sigma^{Z_c 2}_{\upsilon}f^{*2}_{Z_c} e^{-\frac{\mu_{Z_c}^2}{M^2}} +\Sigma^{Z 2}_{\upsilon}f^{*2}_{Z} e^{-\frac{\mu_{Z}^2}{M^2}} & = & \mathbf{ \Upsilon}^{QCD}_{5}(M^2, s_0^*)\label{SR5}.
\end{eqnarray}

The mass sum rules for $Z$ state can be derived by different ways from the above equations. We apply derivative with respect to $-\frac{1}{M^2}$  to both sides of  Eq. (\ref{SR1}) and eliminate $f^{*2}_{Z}$ by dividing both sides of the resultant equation to both sides of the original one. As a result we get 
\begin{equation}\label{muSquare}
\mu_{Z}^2 = \frac{\frac{\partial \mathbf{ \Upsilon}^{QCD}_{1}(M^2, s_0^*)}{\partial\Big(-\frac{1}{M^2}\Big)} -m^{*2}_{Z_c}f^{*2}_{Z_c} \mu_{Z_c}^2 e^{-\frac{\mu_{Z_c}^2}{M^2}} }{\mathbf{ \Upsilon}^{QCD}_{1}(M^2, s_0^*)-m^{*2}_{Z_c}f^{*2}_{Z_c} e^{-\frac{\mu_{Z_c}^2}{M^2}}},
\end{equation}
where, the  in-medium mass of $Z$ state is found as
 \begin{equation}\label{NMmass}
m^{*2}_{Z} = \mu_{Z}^2 + \Sigma^{Z 2}_{\upsilon} - 2p_0 \Sigma^{Z}_{\upsilon}.
\end{equation}
By rearranging the equations  (\ref{SR5}) and   (\ref{SR2})  and dividing both sides of these equations to each other  we get $\Sigma^{Z 2}_{\upsilon}$ as 
\begin{equation}\label{sigNuZ}
\Sigma^{Z 2}_{\upsilon} = \frac{ \mathbf{ \Upsilon}^{QCD}_{5}(M^2, s_0^*) -\Sigma^{Z_c 2}_{\upsilon}f^{*2}_{Z_c} e^{-\frac{\mu_{Z_c}^2}{M^2}} }{\mathbf{ \Upsilon}^{QCD}_{2}(M^2, s_0^*)-f^{*2}_{Z_c} e^{-\frac{\mu_{Z_c}^2}{M^2}}}.
\end{equation}
To obtain the in-medium current coupling constant, $f^{*}_{Z}$, there are different possibilities. One way is to find it from  Eq. (\ref{SR1}), which leads to
\begin{equation}\label{fZstar}
f_{Z}^{*2} = \frac{ \mathbf{ \Upsilon}^{QCD}_{1}(M^2, s_0^*) -m^{*2}_{Z_c}f^{*2}_{Z_c} e^{-\frac{\mu_{Z_c}^2}{M^2}} }{m^{*2}_{Z} e^{-\frac{\mu_{Z}^2}{M^2}}}.
\end{equation}
As it is seen, the parameters of $ Z_c $ calculated in Ref. \cite{Azizi:2020itk} are entered as the inputs to the sum rules for $ Z $ state in the dense medium.

\section{Numerical results}
The expressions obtained in  Eqs. (\ref{muSquare}-\ref{fZstar}) for the physical observables in cold nuclear matter contain quark and baryon masses, nuclear matter density and  different operators both in vacuum and  dense medium as  input parameters. 
The operators appearing in the calculations in a dense medium are presented in Ref. \cite{Azizi:2014yea}. The numerical results for expectation values of different operators  together with the values of other input parameters are collected in table (\ref{tab1}).
\begin{table}[ht!]
\centering
\begin{tabular}{|c|c|c|c|}
\hline  
Parameter & Value & Unit &Ref. \# \\ 
\hline 
$\rho^{sat}$ & $0.11^3$ & GeV$^3$ & \cite{PhysRevC.45.1881} \\
$p_0$ & $4.478^{+0.015}_{-0.018}$ & GeV & \cite{PhysRevD.98.030001} \\
$m_u$ & $2.16^{+0.49}_{-0.26}$ & MeV & \cite{PhysRevD.98.030001} \\ 
$m_d$ & $4.67^{+0.48}_{-0.17}$ & MeV & \cite{PhysRevD.98.030001} \\ 
$m_c$ & $1.27\pm0.02$ & GeV & \cite{PhysRevD.98.030001} \\ 
$\langle q^{\dagger}q\rangle_{\rho}$ & $\frac{3}{2}\rho$ & GeV$^3$ & \cite{COHEN1995221} \\
$\langle \bar{q}q\rangle_{0}$ & $-[272(5)]^3$ & MeV$^3$ & \cite{GUBLER20191,Bazavov:2010yq,PhysRevD.88.014513,PhysRevD.90.114504,PhysRevD.93.054502,10.1093/ptep/ptw129} \\
$\langle \bar{q}q\rangle_{\rho}$ & $\langle \bar{q}q\rangle_{0}+\frac{\sigma_{\pi N}}{2 m_q} \rho$ & MeV$^3$ & \cite{PhysRevC.47.2882} \\
$m_q$ & $0.00345$ & GeV &\cite{PhysRevD.98.030001} \\
$\langle \frac{\alpha_s}{\pi}G^2\rangle_{0}$ & $0.012\pm0.004$ & GeV$^4$ &\cite{GUBLER20191,SHIFMAN1979385,SHIFMAN1979448} \\
$\langle \frac{\alpha_s}{\pi}G^2\rangle_{\rho}$ & $\langle \frac{\alpha_s}{\pi}G^2\rangle_{0} +\rho \langle \frac{\alpha_s}{\pi}G^2 \rangle_N$& GeV$^4$ &\cite{GUBLER20191,Mishra1993,Sait1998} \\
$ \langle \frac{\alpha_s}{\pi}G^2 \rangle_N$ & $ -\frac{8}{9}(M_N-\sigma_{\pi N}-\sigma_{s N}) $& GeV &\cite{GUBLER20191,PhysRevC.45.1881} \\
$M_N$ & $939.49\pm0.05$ & MeV & \cite{PhysRevD.98.030001} \\
$\sigma_{\pi N}$ & $45(6)$ & MeV & \cite{PhysRevD.87.074503} \\
$\sigma_{s N}$ & $21(6)$ & MeV & \cite{PhysRevD.87.074503} \\
$\langle q^{\dagger}iD_0q \rangle_{\rho}$ & $\frac{3}{2}m_q \rho$ & GeV$^4$ &\cite{GUBLER20191,PhysRevC.47.2882} \\
$\langle q^{\dagger}g_s\sigma Gq \rangle_{\rho}$ & $m^2_0\langle q^{\dagger}q\rangle_{\rho}$ & GeV$^5$ &\cite{GUBLER20191} \\
$\langle  \bar{q}g_s\sigma Gq \rangle_{0}$ & $m^2_0 \langle \bar{q}q\rangle_{0}$ & GeV$^5$ &\cite{GUBLER20191} \\
$m^2_0$ & $0.8 \pm 0.2$ & GeV$^2$ & \cite{GUBLER20191,Belyaev:1982sa} \\
$\langle  \bar{q}g_s\sigma Gq \rangle_{\rho}$ & $\langle  \bar{q}g_s\sigma Gq \rangle_{0} +\rho \frac{m^2_0\sigma_{\pi N}}{2m_q}$ & GeV$^5$ &\cite{GUBLER20191,PhysRevC.47.2882} \\
$\langle  \bar{q}iD_0iD_0q \rangle_{\rho}$ & $0.3 ~GeV^2\rho-\frac{1}{8}\langle  \bar{q}g_s\sigma Gq \rangle_{\rho}$ & GeV$^5$ &\cite{GUBLER20191,PhysRevC.47.2882} \\
 \hline 
 \end{tabular}
\caption{Input parameters and their numerical values used in calculations.}\label{tab1}
\end{table}

The obtained in-medium sum rules  for the  mass, current coupling and vector self-energy of $Z$ state contain two more auxiliary parameters, as well. These parameters are fixed based on the standard requirements of the method: mild variations of the physical quantities with respect to the changes in the values of these parameters, dominance of first two resonances over the higher states and continuum and convergence of the series of the  operator product expansion (OPE). 
 To this end, we introduce the quantities
\begin{equation}
\mbox{FTRC}=\frac{\Upsilon_i^{QCD} (M^{2},s^*_{0})}{\Upsilon_i^{QCD} (M^{2},\infty )},  \label{eq:PC}
\end{equation}
where $\mathrm{FTRC} $ stands for first two resonance contribution, and
\begin{equation}
R=\frac{\Upsilon_i ^{\mathrm{Dim5}}(M^{2},s^*_{0})}{\Upsilon_i^{QCD} (M^{2},s^*_{0})}.
\label{eq:Convergence}
\end{equation}%
In Eq.\ (\ref{eq:Convergence}) $\Upsilon_i^{\mathrm{Dim5}}(M^{2},s_{0})$ is the sum of all operators with mass dimension 5, the highest dimension that we consider in the calculations.  The $\mbox{FTRC} $ is used to fix upper bound for $M^{2}$, whereas $R(M^{2})$
is necessary to fix the  lower limit for the Borel parameter. To fix the working regions, we impose $\mbox{FTRC} \geq 0.5$  and $ R > 0.1 $ in order to guarantee the first two resonances dominance  over the higher states and continuum and the OPE convergence.  
 Our analyses show that  all of the above requirements are satisfied when  
\begin{eqnarray}
\centering
\left\{\begin{array}{ll}
M^2& \mbox{in} ~  [3 - 5] ~ \textrm{GeV}^2, \\
s_0^{*} & \mbox{in} ~ [21.9 -23.8] ~ \textrm{GeV}^2.
\end{array}\right.
\end{eqnarray}
\begin{figure}[h!]
\centering
\begin{tabular}{c}
\epsfig{file=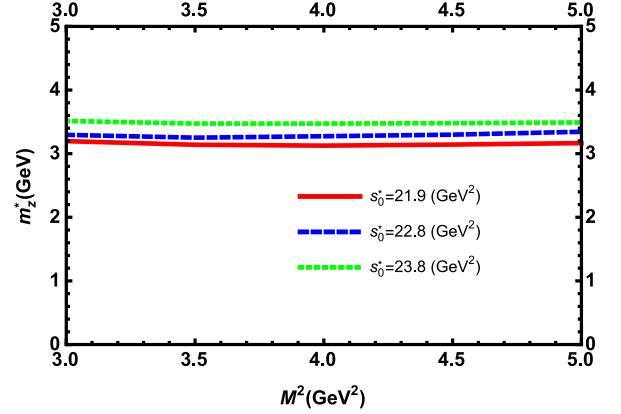,width=0.90\linewidth,clip=}  
\end{tabular}
\caption{The in-medium mass of $Z$ state as functions of Borel mass parameter $M^2$ at the nuclear matter saturation density and at fixed values of the continuum threshold.}\label{Fig1}
\end{figure}
\begin{figure}[h!]
\centering
\begin{tabular}{c}
\epsfig{file=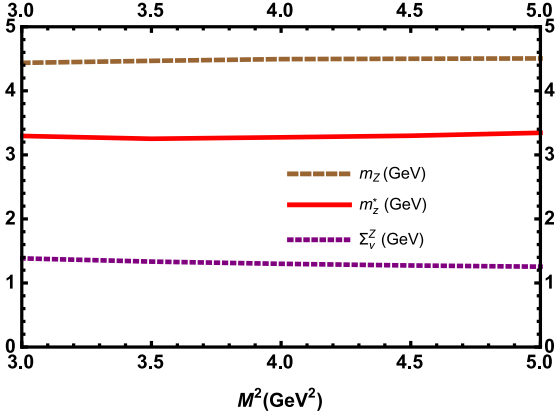,width=0.90\linewidth,clip=}  
\end{tabular}
\caption{The vacuum mass, in-medium mass and vector self-energy of $Z$ state as functions of $M^2$ at average value of continuum threshold and at nuclear matter saturation density.}\label{Fig2}
\end{figure}
\begin{figure}[h!]
\centering
\begin{tabular}{c}
\epsfig{file=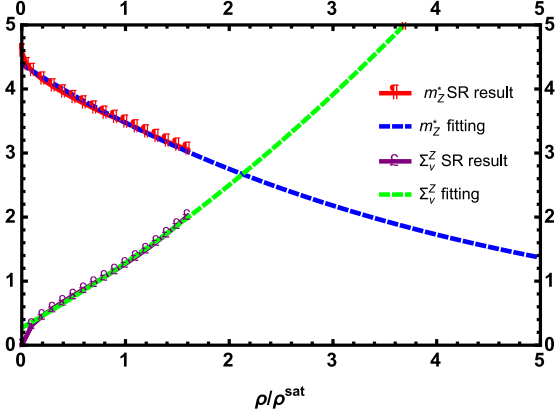,width=0.90\linewidth,clip=}  
\end{tabular}
\caption{In-medium mass and vector self-energy as functions of $\rho/\rho^{sat}$ at mean value of the continuum threshold and Borel parameter.}\label{Fig3}
\end{figure}
\begin{figure}[h!]
\centering
\begin{tabular}{c}
\epsfig{file=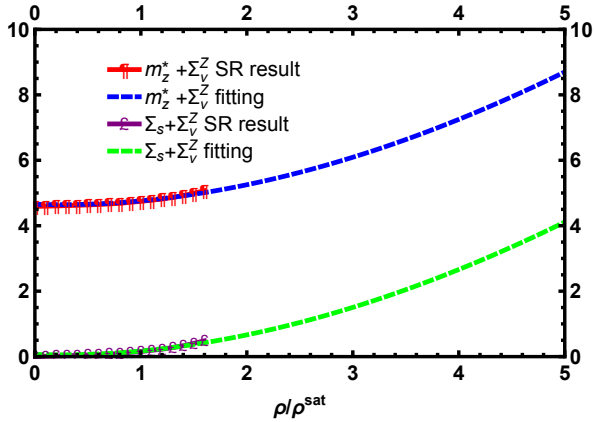,width=0.90\linewidth,clip=}  
\end{tabular}
\caption{The $\Sigma_s + \Sigma^Z_{\upsilon}$ and  $m_Z^* + \Sigma^Z_{\upsilon}$ as  functions of $\rho/\rho^{sat}$ at mean value of the continuum threshold and Borel parameter.}\label{Fig44}
\end{figure}
\begin{figure}[h!]
\centering
\begin{tabular}{c}
\epsfig{file=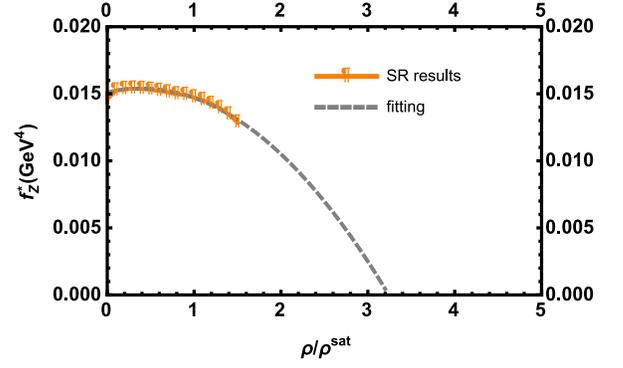,width=0.90\linewidth,clip=}  
\end{tabular}
\caption{The current coupling of $Z$ state as a function of $\rho/\rho^{sat}$ at mean value of the continuum threshold and Borel parameter.}\label{Fig4}
\end{figure}
\begin{figure}[h!]
\centering
\begin{tabular}{c}
\epsfig{file=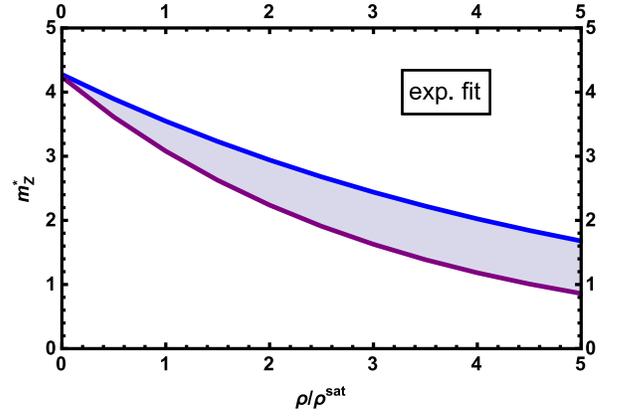,width=0.90\linewidth,clip=}  
\end{tabular}
\caption{The mass $m^*_Z$ as a function of density when the uncertainties of the auxiliary and other input parameters are considered.}\label{Fig5}
\end{figure}
\begin{figure}[h!]
\label{fig1}
\centering
\begin{tabular}{ccc}
\epsfig{file=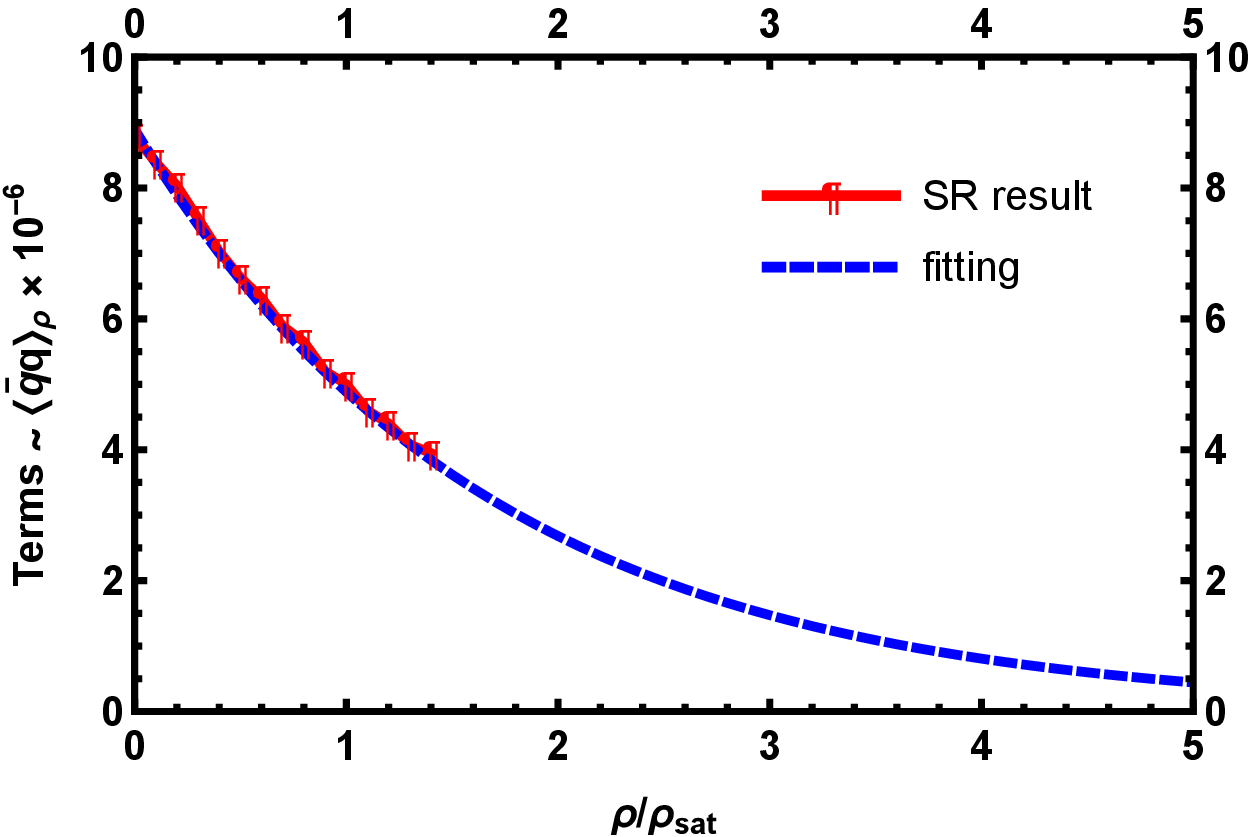,width=0.45\linewidth,clip=} &
\epsfig{file=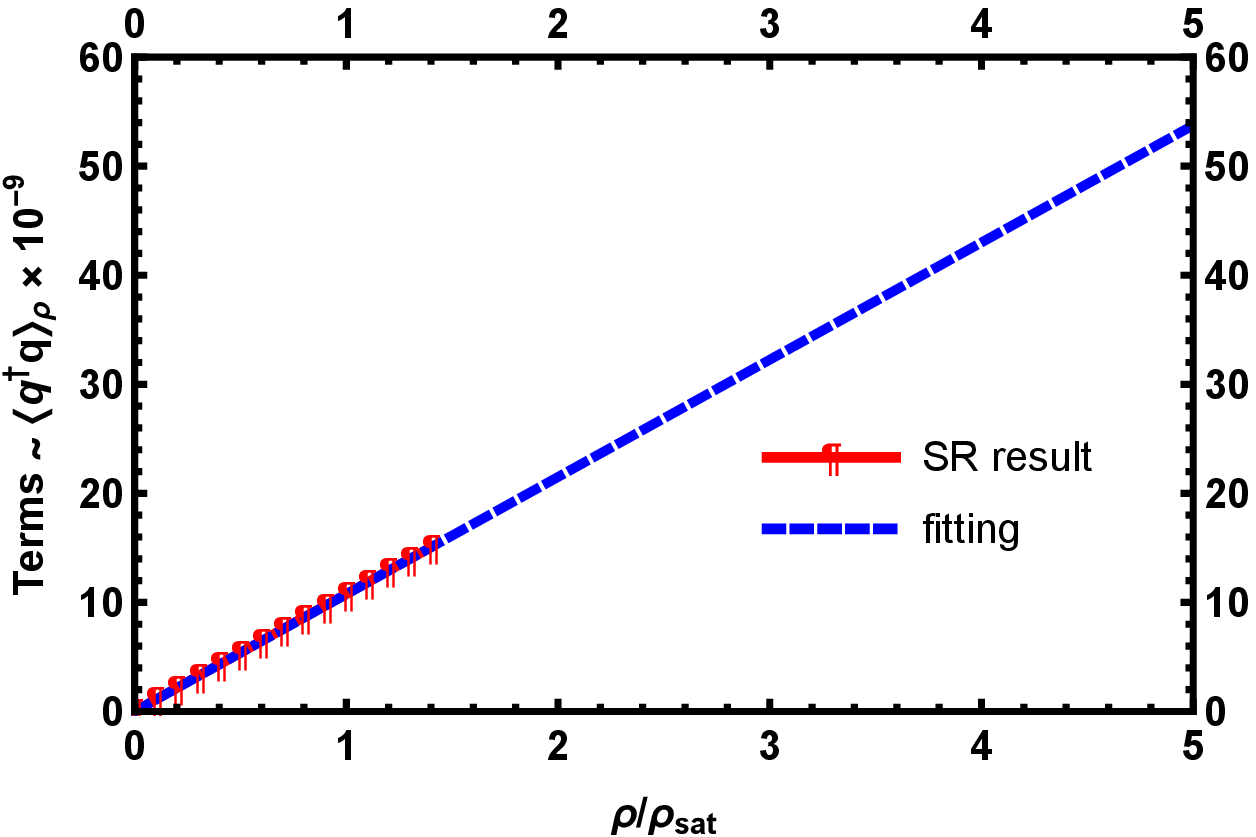,width=0.45\linewidth,clip=}  \\
\epsfig{file=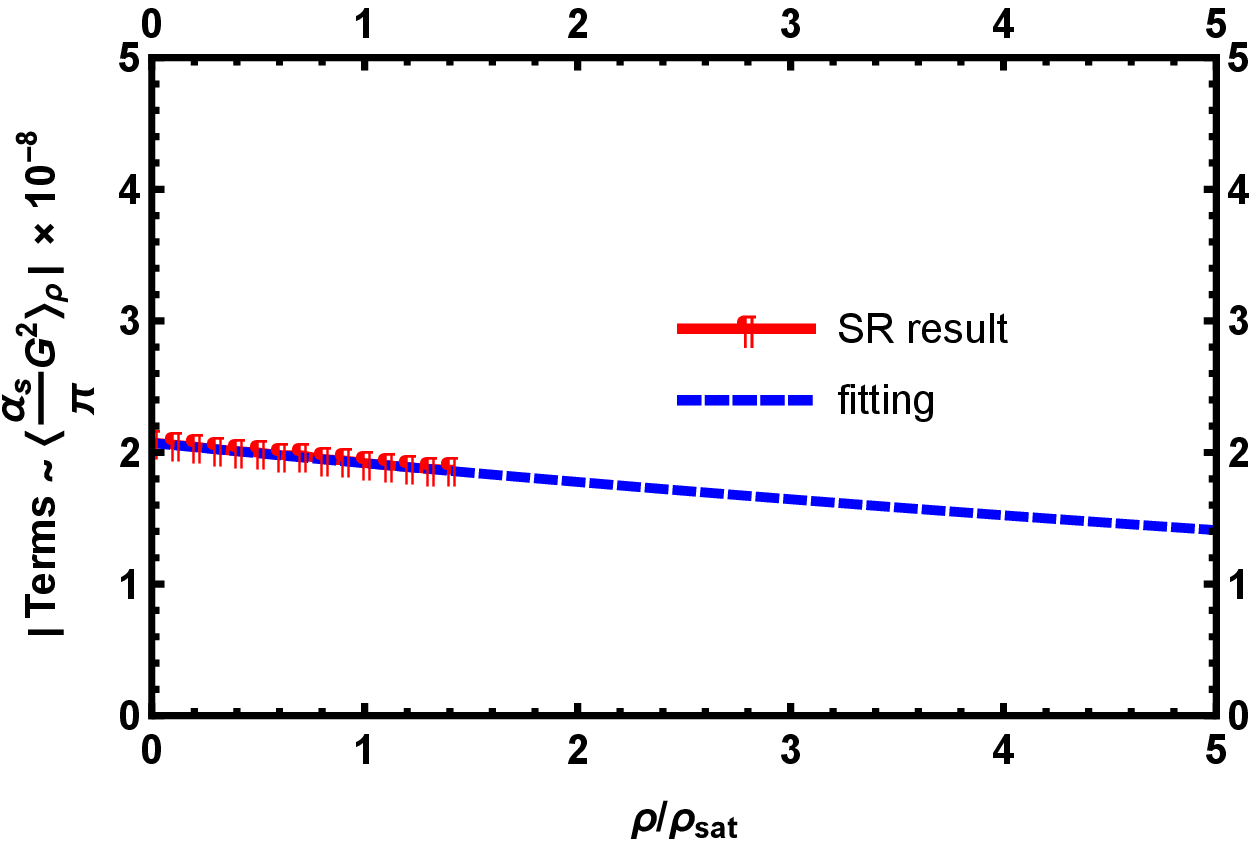,width=0.45\linewidth,clip=} &
\epsfig{file=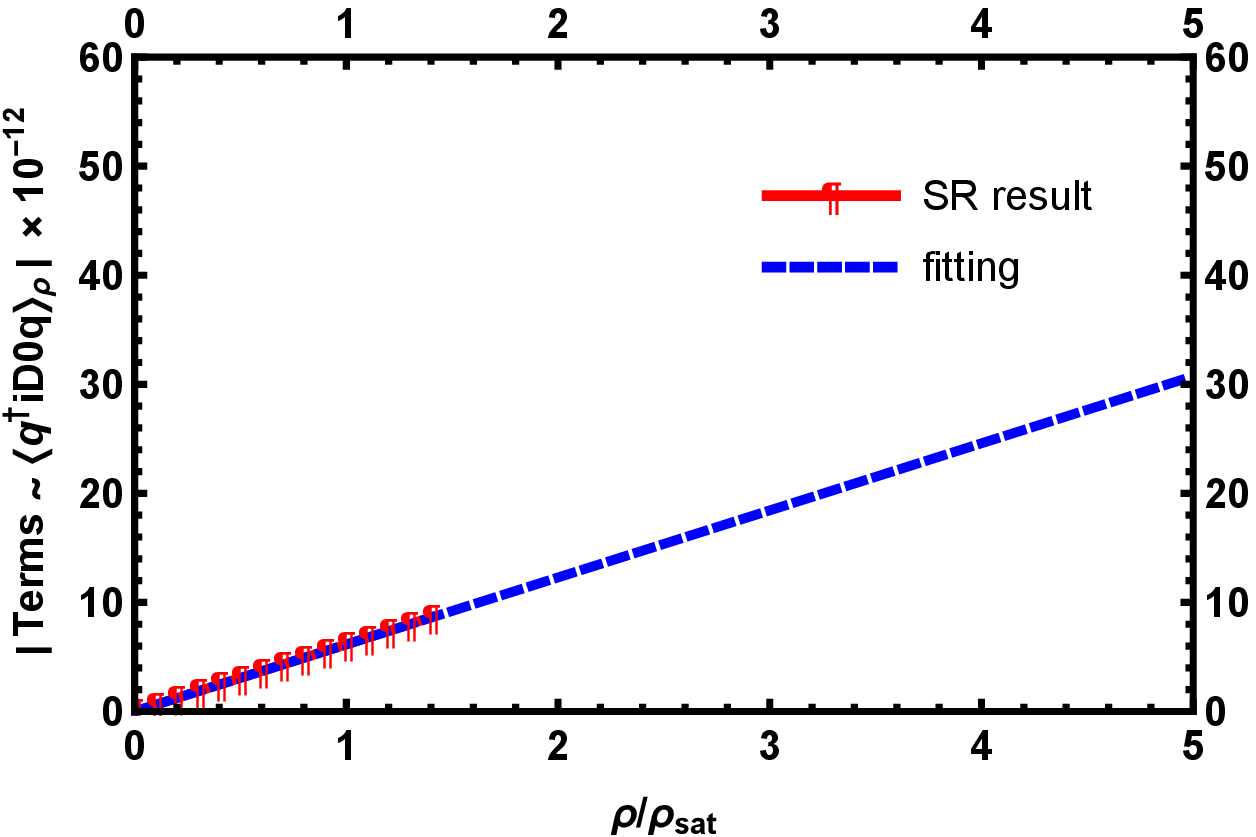,width=0.45\linewidth,clip=}
  \\
\epsfig{file=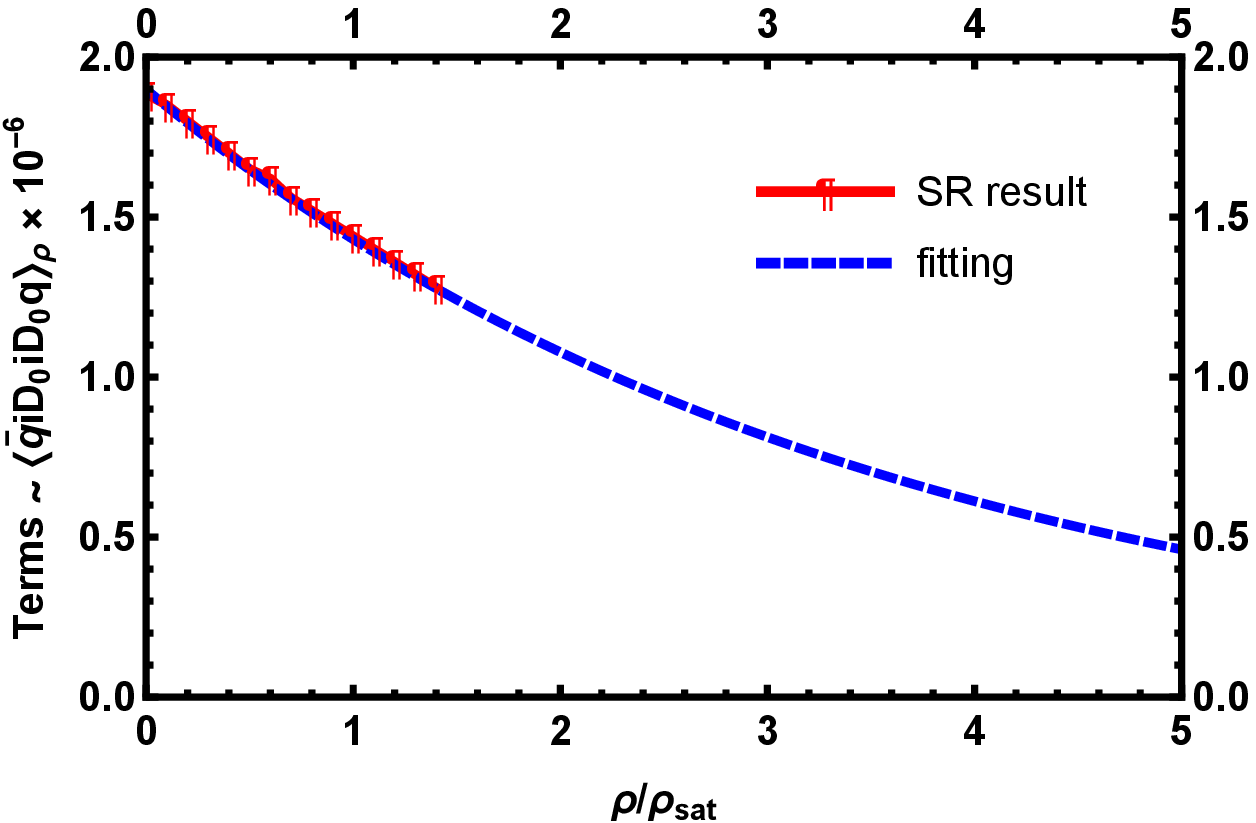,width=0.45\linewidth,clip=} &
\epsfig{file=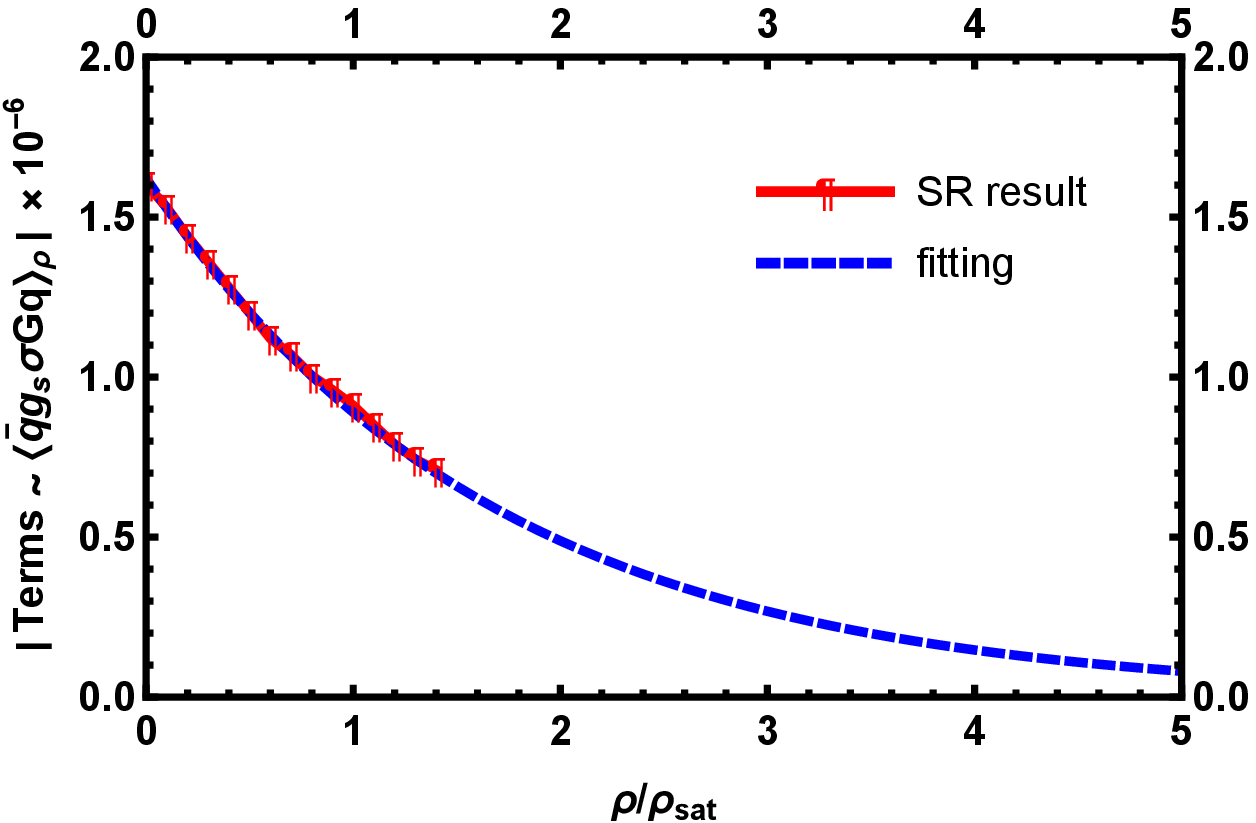,width=0.45\linewidth,clip=}
\end{tabular}
\caption{The dependence of terms proportional to various condensates, entering the QCD side of the calculations, on $\rho/\rho^{sat}$ at average values  of the auxiliary parameters. }\label{Fig7}
\end{figure}

\begin{widetext}
\begin{table}[ht!]
\centering
\begin{tabular}{l|c|c|c|c}
\hline\hline
  & $m_{Z}$  & $m^*_{Z}$   & $f_{Z}.10^2$  & $f^*_{Z}.10^2$    \\
 \hline
PS & $4486^{+112}_{-113}$ MeV & $3311^{+203}_{-185}$ MeV &$1.27^{+0.53}_{-0.36}$ GeV$^4$ & $1.34^{+0.78}_{-0.45}$ GeV$^4$ \\
Exp.  \cite{PhysRevD.98.030001}& $4478^{+15}_{-18}$ MeV  & - & - & - \\
\cite{PhysRevD.96.034026} & $4452^{+182}_{-228}$ MeV & - & $1.48^{+0.31}_{-0.42}$ GeV$^4$ & - \\ 
\cite{Wang:2014vha} & $4.51^{+0.17}_{-0.09}$ GeV &- & $ (5.75^{+98}_{-78})/m_{Z}$ GeV$^4$ &- \\
\cite{Lee:2007gs} & $4.40 \pm 0.10$ GeV & - & -  &- \\
\cite{Bracco:2008jj} & $4.52 \pm 0.09$ GeV & - & $(3.75 \pm 0.48)/m_{Z}$ GeV$^4$  &- \\
\cite{Chen:2019osl} & $4.53^{+0.16}_{-0.10}$ GeV & - & -  &- \\
\hline\hline
\end{tabular}
\caption{The vacuum and in-medium masses and current couplings of $Z$ state. PS stands for present study.}\label{tab2}
\end{table}
\end{widetext}
 
In  Fig. (\ref{Fig1}), we demonstrate the in-medium mass $m^*_Z$ as a function of Borel mass parameter $M^2$ at the nuclear matter saturation density, $\rho^{sat}=0.11^3$ GeV$^3$, and at three fixed values of the continuum threshold. As is seen, the in-medium mass of the $Z(4430)$ resonance shows pretty good stability against variations of both of Borel mass parameter $M^2$ and in-medium continuum threshold $s_0^{*}$. In  Fig. (\ref{Fig2}), the vacuum mass ($m_Z$) which is obtained  at $\rho \rightarrow 0$ limit, the  in-medium mass ($m^*_Z$) and vector self-energy  ($\Sigma^Z_{\upsilon}$) of $Z$ state are displayed as a function of $M^2$ at average value of continuum threshold and at nuclear matter saturation density. From this figure we see a considerable negative shift in the mass of the state under study due to the medium effects. This shit is called the scalar self-energy.  Strictly speaking,  the in-medium mass of the $Z$ resonance reduces to approximately $74\%$ of its vacuum value at the nuclear matter saturation density and amounts as $3311^{+203}_{-185}$ MeV. Thus, the negative shift in the mass due to the cold nuclear matter is approximately $26\%$. This state gains a  vector self-energy  with the  value $1310^{+101}_{-86}$ MeV at nuclear matter saturation density. At saturation density the current coupling of the state $Z$ is  found as $1.34^{+0.78}_{-0.45}$ GeV$^4$.
The numerical  results  for the masses and current couplings  obtained in vacuum and medium are collected in table (\ref{tab2}). For comparison, we demonstrate the existing values from the experiment and other theoretical studies in the same table, as well.   From this table, we see that the vacuum mass obtained in the present study is nicely consistent with the world experimental  average \cite{PhysRevD.98.030001}. This value is also in accord with other theoretical predictions within the errors. The vacuum current coupling is consistent with other theoretical predictions within the presented uncertainties. Our predictions for the in-medium mass and current coupling as well as the vector self-energy can be tested in future in-medium experiments and by other theoretical studies. 

The main aim in the present study, is to obtain the behavior of the physical quantities under study with respect to the density, specially at higher densities. In the present study, we consider the range from zero to $ 5\rho^{sat} $, corresponding to the density of the  cores of massive neutron stars. Neutron stars, as natural QCD matter laboratories,  are very dense and compact  that may produce strange and heavy baryons as well as the  exotic states  regarding the processes that may occur in their inside.  For the neutron stars with mass $ \sim1.5M_{\bigodot}$,   the corresponding core density is approximately (2-3) $\rho^{sat} $ and for the mass  $ \geq 2M_{\bigodot}$  the  density is about $ 5\rho^{sat} $ of the nuclear matter \cite{KKim}. To understand the process occurring at  higher densities with the aim of getting knowledge on the internal structures of the neutron stars as well as analyzing of the future data in the heavy ion collision experiments, we need to investigate the properties of various hadrons under extreme conditions. In the present study, we deal with cold nuclear matter ($ T\simeq0 $), hence there is no phase transition expected up to the density that we consider in the calculations. The phase transition to QGP is expected to occur at points  $\geq 5.8 \rho^{sat} $ at zero temperature \cite{Alice}. The QGP state of matter is studied in current heavy ion collision experiments at the LHC/CERN and   RHIC at BNL. Presently,  the QCD phase diagram of quark matter is not well known, either experimentally or theoretically, but there are some information from the lattice QCD  and other phenomenological models. Many experiments are designed with the aim of investigation of the hadronic properties and possible phase transitions at finite temperatures and densities.  As previously mentioned, the FAIR at GSI and the NICA at Dubna aim to provide useful information in this respect. 

To discuss the dependence of the physical quantities on the density, 
 in Fig. (\ref{Fig3}) we display $m^*_Z$ and $\Sigma^{Z }_{\upsilon}$ as  functions of $\rho/\rho^{sat}$ at average values of the continuum threshold and Borel parameter. As it is clear, the sum rules for both the in-medium mass and vector self-energy give reliable results up to $\rho=1.6 \rho^{sat}$ GeV$^3$. To extend our calculations for higher densities, we use some fit functions. The best fit for the mass is the following exponential function:
\begin{equation}
\label{ }
m^*_Z (\rho) = 4.384 e^{-0.233x} ~ \textrm{GeV},
\end{equation}
where $ x=\rho/\rho^{sat} $. For vector self-energy we obtain
\begin{equation}
\label{ }
\Sigma^Z_{\upsilon} (\rho) =0.099x^2 + 0.917x + 0.272 ~ \textrm{GeV}.
\end{equation}
In order to better understand how $Z(4430)$ feels the total potential, and how much energy would be necessary to produce this state in experiments using nuclear targets, in Fig.  (\ref{Fig44}), we show the dependence of $\Sigma_s + \Sigma^Z_{\upsilon}$, or $m_Z^* + \Sigma^Z_{\upsilon}$ on $\rho/\rho^{sat}  $. We observe that both of these quantities grow with increasing in the density of the medium.

In Fig. (\ref{Fig4}), we plot the density-dependent current coupling. The following function best describes the current coupling:
\begin{equation}
f^* _Z (\rho)=-0.002x^2 + 0.002x + 0.015  \textrm{GeV}^4.
\end{equation}
The current coupling becomes zero at $ x=3.2 $. The vector self-energy increases with the increasing in the density, while the mass decreases with increasing in the density and reaches to roughly $ 1.4GeV $ at $ x=5 $ in average.  Fig.  (\ref{Fig5}) shows the $m^*_Z-x  $ graphic when all the uncertainties in the auxiliary parameters as well as in other inputs are considered.  As it is clear from the presented figures, the parameters of the $ Z(4430) $ state depend on the density,  considerably.   In \cite{Azizi:2014yea,COHEN1995221,Navarra:2004bj,Azizi:2015ica}, which study the behavior of the nucleon, hyperon and light $ \Theta^+ $ pentaquark with respect to density up to  $\rho\simeq1.5\rho^{sat}$, the dependence of the parameters on density are relatively small.  In the present study we consider to investigate the behavior of many parameters of the doubly-heavy excited $ Z(4430) $ tetraquark state up to $5\rho^{sat}$, corresponding to the density of the cores of massive neutron stars, and observe that the considered parameters  depend on the density, remarkably. These behavior can be attributed to the considerable dependence of the  various terms proportional to different condensates in the QCD side of the present calculations on the density as depicted in Fig. (\ref{Fig7}).

\section{Summary and conclusions}
The vacuum properties of different tetraquark states have been discussed in many theoretical studies as well as different experiments. With the developments in the experimental side, we hope we will be able to study the behavior of such  states in in-medium experiments in near future. Thus, theoretical studies on the spectroscopic properties of exotics can play crucial roles in conducting the related experiments. Such studies can also help us fix the quantum numbers and determine the nature and internal structures of the famous candidates of charmonium-like tetraquarks.  In this accordance, we took into consideration the $ Z(4430) $ state in the present study and calculated some spectroscopic parameters of this state both in vacuum and a medium with finite density.  Based on some experimental information, we treated this state as the first excited state in the $ Z_c(3900) $ channel and constructed the related sum rules based on the standard prescriptions of the method. As we saw, the parameters of the ground state $ Z_c(3900) $  were entered as input parameters to the sum rules for $ Z(4430) $ state.  We fixed the entering auxiliary parameters in accordance with the requirements of the method and used the expectation values of the different in-medium operators available in the literature to extract the values of the parameters at saturation density and determine the behavior of these parameters with respect to the density. 

The numerical calculations showed that the mass gains a negative shift (scalar self-energy) due to the medium effects: this shifts amounts $ -26\% $ at saturation density. This state gains a repulsive vector self-energy which amounts  $1310^{+101}_{-86}$ MeV at saturation density. We also found the values of the current coupling as  $1.34^{+0.78}_{-0.45}$ GeV$^4$ at nuclear matter saturation density. This value can be used in determinations of different parameters related to the electromagnetic, weak and strong interactions/decays of this state in medium. 

 We obtained the behavior of the physical quantities in the interval $ \rho\in [0,5] \rho^{sat}$, which its upper limit corresponds to the core density of the neutron stars. We observed that the current coupling becomes zero at $\rho=3.2 \rho^{sat}$, while the vector self-energy increases with the density. The mass exponentially reduces with the increasing in the value of the density and reaches to roughly $ 1.4GeV $ at the end point. With the progress in the experimental side we hope that the heavy ion collision experiments at CERN and BNL as well as the FAIR at GSI and  NICA at Dubna will be able to provide useful information on the behavior of the hadronic matter under extreme conditions as well as possible phase traditions to QGP. With upgraded detectors at RHIC and LHC, it has become possible to measure the properties of  hadrons beyond their ground states. Thus, these experiments provide a new platform to experimentally study the exotic hadrons that are either in molecular structure made of standard  hadrons or compact objects of multiquarks and antiquarks.
 The  $ \bar{P}ANDA $ experiment at FAIR is also expected to investigate the properties of the standard  hadrons as well as the exotic states like tetraquarks, including the $ Z_c(3900) $ and $ Z(4430) $ states, at finite density. 
 

\section*{Acknowledgments}

The authors thank  TUBITAK for the  partial support provided under the Grant No. 119F094.


\end{document}